\documentstyle[aps,prb,preprint,tighten]{revtex}

\begin{document}
\draft
%
\def\rf#1{${}^{#1}$}
%
\def\hbn{\hfil\break\noindent}
\def\hb{\hfil\break}
\def\bn{\break\noindent}
%
\def\gsim{\buildrel >\over\sim }
\def\lsim{\buildrel <\over\sim }
%
\title{Contrasting Dynamic Spin Susceptibility Models\\
and their Relation to High Temperature Superconductivity}
\author{H.-B. Sch\"{u}ttler}
\address{Center for Simulational Physics,
Department of Physics and Astronomy,
The University of Georgia, Athens, Georgia  30602}
\author{M. R. Norman}
\address{Materials Science Division,
Argonne National Laboratory, Argonne, Ilinois  60439}
\maketitle

\begin{abstract}
We compare the normal-state resistivities $\rho$ and
the critical temperatures $T_c$ for superconducting $d_{x^2-y^2}$ pairing
due to antiferromagnetic (AF) spin fluctuation exchange
in the context of the two
phenomenological dynamical spin susceptibility models,
recently proposed by Millis, Monien, and Pines (MMP) and
Monthoux and Pines (MP) and, respectively,
by Radtke, Ullah, Levin, and Norman (RULN), for the cuprate
high-$T_c$ materials.
Assuming comparable electronic bandwidths and resistiviies
in both models, we show that the RULN model gives a much lower
$d$-wave $T_c$ ($\lsim\!20$K) than the MMP model
(with $T_c\!\sim\!100$K).
We demonstrate that these profound differences in the $T_c$'s
arise from fundamental differences in the
spectral weight distributions of the two model susceptibilities
and are {\it{not}} primarily caused by differences in
the calculational techniques
employed by MP and RULN.
The MMP model, claimed to fit NMR data in YBCO,
exhibits substantial amounts of spin fluctuation
spectral weight up to an imposed  cut-off of $400$meV,
whereas, in the RULN model, claimed to fit
YBCO neutron scattering data,
the weight is narrowly peaked and
effectively cut-off by $100$meV.
Further neutron scattering experiments,
to explore the spectral weight distribution at
all wavevectors over a sufficiently large excitation energy range,
will thus be of crucial importance to resolve the question
whether AF spin fluctuation exchange provides
a viable mechanism to account for high-$T_c$ superconductivity.
The large high-frequency boson spectral weight, needed to generate
both a high $d$-wave $T_c$ and a low normal-state resistivity,
also implies large values, of order unity,
for the  Migdal smallness parameter,
thus casting serious doubt on the validity of the very
Migdal approximation on which high-$T_c$ spin fluctuation
exchange models are based.
\end{abstract}

\bigskip


\narrowtext

\hbn
\section{Introduction}
\label{sec:I}

Recently, an increasing, but not uncontroverted
amount of experimental data has been shown to be
consistent with a $d_{x^2-y^2}$ symmetry of the superconducting order
parameter in the cuprate high-$T_c$
superconductors.\cite{phystod,scalapino-review}
In early studies of the Hubbard model,
Scalapino and co-workers\cite{scalapino-loh-hirsch}
proposed that a pairing state of $d$-wave symmetry
can arise due to exchange of
antiferromagnetic spin fluctuations.\cite{miyake-schmitt-rink-varma}
In the model believed to be most relevant to the cuprates,
the two-dimensional (2D) Hubbard model,
a $d_{x^2-y^2}$ pairing instability was suggested by
Bickers {\it et al.}\cite{bickers} and by Moriya {\it et al.}\cite{moriya}
Recently, two phenomenological models of the dynamic
spin susceptibility $\chi(\vec q,\omega)$  for the cuprate materials
have been developed by fitting an assumed analytical
form to experimental data and
then using this information to calculate such properties as
the resistivity $\rho$ and superconducting transition temperature $T_c$.

The first model, due to Millis, Monien, and Pines, hereafter referred to
as  MMP, was obtained from fits to nuclear magnetic resonance (NMR)
data on $YBa_2Cu_3O_7$ ($YBCO$)\cite{mmp}.
This model and modifications thereof were then used to calculate
$T_c$ in weak coupling\cite{mpb} and strong coupling\cite{mp,mp1,mp2}
approaches.  Moreover,
the resistivity due to spin fluctuation scattering was calculated, both
without\cite{mp1} and with\cite{mp2} inclusion of
leading-order vertex corrections.
Also calculated was the suppression of $T_c$
due to non-magnetic impurities.\cite{mp2}
The primary conclusion drawn from the MMP model
calculations\cite{mmp,mpb,mp,mp1,mp2}
was that the NMR data, the temperature-dependent normal-state resistivity,
the large superconducting transition temperature
$T_c\!\cong\!100K$,
{{}{and}} the $T_c$-suppression due to impurities
can be consistently accounted for with a single,
physically acceptable set of model parameters.

The second model, due to Radtke, Levin, Ullah, and Norman,\cite{ruln}
referred to as RULN hereafter, was obtained from fits
to neutron scattering data on $YBCO$.\cite{sizha}
This $\chi(\vec q,\omega)$ was then used to calculate $T_c$, using a
strong coupling approach\cite{ruln,rlsn-vhs}, the resistivity, by
variational Boltzmann equation and
Fermi-surface-restricted Kubo formalisms\cite{ruln},
and the $T_c$ suppression due to impurities.\cite{rlsn-imp}
The primary conclusion drawn from these calculations was
that it is essentially impossible to generate a $d$-wave $T_c$
of $90K$ (or larger) by AF spin fluctuation exchange.
In fact, if the effective spin fluctuation coupling constant
in the RULN model was adjusted to reproduce the observed
magnitude of the normal state resistivity, the resulting
$d$-wave $T_c$ was found not to exceed about $10{\rm{K}}$.

We are thus faced with the obvious controversy
that the two different model fits to the dynamic spin
susceptibility of $YBCO$ give strikingly different results
for the $d$-wave $T_c$ and/or the normal-state transport scattering rates.
The MMP model allows to obtain a very high $T_c$,
consistent with the cuprates, as well as a rather low resistivity,
consistent with recent single crystal data on $YBCO$.
The RULN model, on the other hand, gives a much lower upper
limit on $T_c$, even for resistivities much larger than the MMP model.

This controversy has led to a number of criticisms on both sides.
An early argument\cite{mp1,ruln} concerned the use of different
calculational approaches, specifically the use of
weak-coupling vs. strong-coupling and
approximate semi-analytical Fermi-surface-restricted
vs. exact numerical full-zone approaches to calculate $T_c$.
However, both groups\cite{mp1,mp2,rlsn-vhs}
have now carried out detailed
strong-coupling, full-zone $T_c$ calculations
for {{}{both}} models, using fast Fourier transform (FFT)
techniques\cite{serene-hess} which allow to include
include the full momentum and frequency
dependence of the Migdal-Eliashberg equations.
In these full-zone results, the order-of-magnitude
discrepancy of the $T_c$ values between the two models persists.
Clearly, the different
magnitudes of $T_c$ are {{}{not primarily}} due to the use of
different calculational approaches.

Another objection\cite{mp1,mp2}
against the original RULN approach concerned
their\cite{ruln} use of variational
Boltzmann or
Fermi-surface-restricted Kubo formalisms
for calculating the resistivity in the RULN model.
These transport formalisms, again, amount essentially to an
approximate Fermi-surface-restricted analysis of
the relevant full-zone Bethe-Salpeter integral equation.
By contrast, in Ref.\onlinecite{mp2},
the resistivity in the MMP model was obtained
directly from the numerical full-zone solution of the
Bethe-Salpeter equation, without additional approximations.
In the present paper, we will show that, again,
the discrepancy between the two models is {{}{not}}
primarily caused by the use of two different calculational approaches.

What we wish to demonstrate here is that the
very different conclusions about attainable $T_c$'s
derived from the two models
are a direct consequence of the fundamental differences in the assumed
dynamic spin susceptibilities $\chi(\vec q,\omega)$.
We re-emphasize\cite{rlsn-vhs} that, even if one uses the same electronic
band parameters in both models, the RULN model will still
give a substantially lower superconducting $T_c$
than the MMP model. We will show that this difference in $T_c$
arises, ultimately, from the difference in the spectral
weight distribution of the two model spin susceptibilities.
We therefore conclude that more experimental work is needed, primarily
by neutron scattering, to elucidate the
frequency dependence of the spin susceptibility $\chi(\vec q,\omega)$.
Only a complete understanding of the spectral weight
distribution, including its full frequency {{}{and}}
momentum dependence, will eventually
allow us to resolve the issue of whether or not
a spin fluctuation pairing mechanism
may constitute a viable model of high temperature superconductivity.

On the theoretical side, our analysis of the two different
models shows that, in order to produce a large $T_c$ and,
at the same time, a sufficiently low normal-state resistivity,
the spin fluctuation models, just like any other
boson exchange model, must exhibit substantial amounts
of spectral weight at a fairly large boson energy scale
$\Omega$, as well as a substantial overall
Eliashberg coupling parameter $\lambda$,
assuming typical bandwidths and Fermi energies $\epsilon_F$
in the cuprates. For such large boson energy scales and coupling
strengths, the Migdal smallness parameter
$s\!\equiv\!\lambda\Omega/\epsilon_F$ is not really small anymore,
compared to unity.
Thus, Migdal's "theorem", the very foundation of current
spin fluctuation exchange models, tends to be on rather shaky grounds.

Another important theoretical consequence of having such a large boson
energy scale, $\Omega/\epsilon_F\!\gsim\!0.1-0.2$, is that
the conventional mechanism of evading Coulomb interactions
by retardation becomes largely inoperative in these types of
high-energy boson exchange models.
Thus, even though the local, Hubbard-type on-site Coulomb
correlations are completely projected away,
simply due to the non-$s$-wave order parameter symmetry,
the extended (e.g. nearest neighbor) part of the
Coulomb interaction potential can be quite effective
in suppressing the spin-fluctuation-mediated $d$-wave $T_c$.\cite{hbs-unpub}
Current spin fluctuation models, including the MMP model,\cite{mp,mp1,mp2}
have so far entirely ignored the extended part of the
Coulomb interaction potential and
may thus be severely overestimating the actual $d$-wave $T_c$ values
which can be ultimately achieved via a spin fluctuation exchange
mechanism.

In passing, we should also comment briefly on the on-going
controversy concerning impurity effects. Monthoux and
Pines\cite{mp2} claimed that the
suppression of $T_c$ due to pairbreaking by
non-magnetic impurities is much weaker than would be indicated by
Abrikosov-Gor'kov\cite{ag} (AG) theory.
Radtke {\it et al.}\cite{rlsn-imp} on the other hand claimed
that the suppression of $T_c$ due to non-magnetic impurities
{{}{in both models}} is roughly in agreement
with a strong-coupling version of AG theory proposed by Millis
{\it et al.}\cite{msv}
At this point, we can only emphasize, again,\cite{rlsn-imp}
that the two models are indeed quite consistent in that aspect.
The differences claimed to exist by Monthoux and Pines\cite{mp2}
are entirely a matter of different calculational approaches for
evaluating the impurity scattering rate from the input
impurity potential strength, using either the
$t$-matrix\cite{mp2} or leading-order
Born\cite{rlsn-imp,mp2} approximation.
If the calculated impurity-induced $T_c$-suppression is plotted
as a function of the calculated impurity scattering rate
(rather than as a function of the input impurity potential
strength), then both approaches, Born and $t$-matrix,
give quite similar results. That is, for a given magnitude
of impurity scattering rate, one obtains essentially
the same $T_c$-suppression, regardless of how that
scattering rate was calculated from the input impurity
potential. The important point to remember here
is that it is the impurity scattering rate,
{{}{not}} the input potential strength, which is measured
experimentally via the impurity-induced "residual" resistivity.

The remainder of the paper is organized as follows:

In Section II, we introduce the two AF susceptibility
models, proposed by MMP\cite{mmp} and RULN\cite{ruln},
and summarize the different approaches
used by Monthoux and Pines\cite{mp,mp1,mp2}
(referred to as MP, hereafter)
and by RULN\cite{ruln} in determining
the electronic band parameters and
in calculating transport properties,
for their respective models.

In Section III, we discuss in some detail the
underlying physical assumptions on which, implicitly,
the different approaches for estimating the band parameters
and for calculating transport properties are based.
We then give a direct comparison of the different transport
formalisms, for the case of the MMP model,
and discuss the limits of applicability
of both the Fermi-surface-restricted Boltzmann
and the full-zone Bethe-Salpeter formalisms, and
of the underlying Migdal approximation.

In Section IV, we discuss the origin of the differences in
the $d$-wave pairing $T_c$'s extracted from the two models.
We show that the substantial differences in the
$T_c$'s of the two models are, ultimately, {{}{not}} due to
the differences in the assumed bandwidth parameters or in the
transport formalism used to estimate the coupling parameter.
By means of a McMillan-Allen-Dynes analysis,\cite{ad,allen-mitro}
we demonstrate that the primary cause for the differences in $T_c$
lies in the different spectral weight distributions
of the two model spin susceptibilities.

Concluding remarks are presented in Section V.
\hbn
\hbn
\hbn
\section{Model susceptibilities, band and coupling parameters}
\label{sec:II}

We start by describing the two models. The MMP model is a fit
designed to describe NMR data in $YBCO$, based on a susceptibility
spectral function of the analytical form
\begin{equation}
{\rm{Im}}\chi_{MMP}(\vec q,\omega+i0^+)=
\frac{\chi_Q\omega/\omega_{sf}}
{(1+\xi^2|\vec q - \vec Q|^2)^2+(\omega/\omega_{sf})^2}
\Theta(\Omega_c-|\omega|)
\label{eq:1}
\end{equation}
where $\vec Q = (\pi/a,\pi/a)$ and $\vec q\!\equiv\!(q_x,q_y)$
is in the 1st quadrant of the two-dimensional (2D) square lattice
Brillouin zone and a frequency cut-off taken as $\Omega_c=400{\rm{meV}}$.
In the most recent version of the model,
proposed by Monthoux and Pines\cite{mp2}, hereafter referred
to as MP-II,
the other parameter values are, as given in Table II
of that paper\cite{mp2} for a 25\% hole doping concentration,
\begin{equation}
\omega_{sf}=14{\rm{meV}},\;\;\;\;
\xi/a=2.3,\;\;\;\;
\chi_Q=75 {\rm{eV}}^{-1}.
\label{eq:2}
\end{equation}
where $a$ ($\cong\!3.8\AA$) denotes the 2D $CuO_2$ square lattice constant.
Note, that in MP-II\cite{mp2}, the resistivity
as a function of temperature $T$ was actually calculated by
using a $T$-dependent $\omega_{sf}$, $\chi_Q$, $\xi$
as defined in Eqs. (37), (27) and (5b)  of that paper,\cite{thanks}
respectively, that is, in the notation of MP-II\cite{mp2}
\begin{eqnarray}
\omega_{sf}(T)&=&9.5{\rm{meV}}+4.75\times10^{-2}{\rm{(meV/K)}}\times T
\nonumber\\
\chi_Q(T)&=&\frac{\chi_0\Gamma_{sf}}{\pi\omega_{sf}(T)}
\nonumber\\
\xi(T)/a&=&\Big[\frac{\Gamma_{sf}}{\pi\beta^{1/2}\omega_{sf}(T)}\Big]^{1/2}
\label{eq:3}
\end{eqnarray}
with $\Gamma_{sf}\!=\!1.3{\rm{eV}}$, $\chi_0\!=\!2.6{\rm{eV}}^{-1}$
and $\beta\!=\!32$. At $T\!\cong\!100K$,
these $T$-dependent values Eq.(3)
agree to within a few percent with those of Eq.(2), taken from MP-II's
Table II. In the following we will use exclusively the $T$-dependent
values of $\omega_{sf}$, $\xi/a$, and $\chi_Q$ from Eq.(3).

The RULN model susceptibility, designed to fit
neutron scattering data in $YBCO$, is given by\cite{ruln}
\begin{eqnarray}
{\rm{Im}}\chi_{RULN}(\vec q,\omega+i0^+)
&=& C\Big[\frac{1}{1+J_0(cos(q_xa)+cos(q_ya))}\Big]^2
\nonumber\\
&\times&
\frac{3(T+5)\omega}{1.05\omega^2-60|\omega|+900+3(T+5)^2}
\Theta(\Omega_c-|\omega|)
\label{eq:4}
\end{eqnarray}
with the temperature T and excitation energy $\omega$
measured in meV, a cut-off $\Omega_c\!=\!100$meV,
a constant $J_0\!=\!0.3$ and a prefactor $C=0.19{\rm{eV}}^{-1}$.

In either model, the spin-fluctuation-mediated electron-electron
interaction potential $V(\vec q,\omega)$
is then obtained by multiplying $\chi(\vec q,\omega)$ by a coupling constant
$g^2$, that is
\begin{equation}
V(\vec q,\omega)=g^2\chi(\vec q,\omega)=
g^2\int_{-\infty}^{+\infty} \frac{d\omega'}{\pi}
\frac{{\rm{Im}}\chi(\vec q,\omega'+i0^+)}{\omega'-\omega}
\label{eq:5}
\end{equation}
for transfer of momentum $\vec q$ and complex frequency $\omega$.
To calculate superconducting $T_c$'s and normal state conductivities,
using standard diagrammatic and/or transport theory approaches,
this interaction potential is combined with a
2D tight-binding electron bandstructure
\begin{equation}
\epsilon_{\vec k}\!=\!-2t_1[\cos(ak_x)+\cos(ak_y)
                     -2t\cos(ak_x)\cos(ak_y)-e]
\label{eq:6}
\end{equation}
with a chemical potential $\mu\!\equiv\!2e\!\times\!t_1$
and 1st and 2nd neighbor hopping matrix elements
$t_1$ and $t_2\!\equiv\!t\!\times\!t_1$, respectively.
To fix the ratios $t\!\equiv\!t_2/t_1$ and $e\!\equiv\!\mu/2t_1$,
both RULN\cite{ruln} and, subsequently, MP-II\cite{mp2}
used the Fermi surface shapes and Fermi surface
volumes found in LDA bandstructure
calculations\cite{lda},
which are consistent with angle-resolved
photoemission experiments\cite{arpes}.
In the case of $YBa_2Cu_3O_{6.7}$, considered by RULN,
this gave $t\!\cong\!+0.45$, working in the hole picture,
and a $\mu$ value which corresponds
to a hole doping concentration, measured from half-filling,
of $x\!=\!0.18$ holes per $Cu$-site, that is,
a total hole concentration of $n_h\!\equiv\!1+x\!=\!1.18$ holes
per $Cu$-site.
MP-II adopted the same value for $t$, but assumed a somewhat
larger hole doping concentration of $x\!=\!0.25$
holes per $Cu$-site,
to correspond to the case of $YBa_2Cu_3O_{6.9}$.
However, MP\cite{mp1,mp2} and RULN\cite{ruln} then
employed two very different approaches to determine
the bandwidth $8t_1$ and the spin fluctuation coupling
constant $g^2$:

MP\cite{mp,mp1,mp2} adjusted $t_1$ to match roughly
the calculated LDA bandstructure $\epsilon_{\vec k}$ near the Fermi
surface, giving a $t_1\!\cong\!250$meV for $YBCO$.
They then adjusted their $g^2$ to give a superconducting
$d$-wave $T_c$ of about $100$K, as obtained from
solutions of the full-zone $\vec k$- and $\omega$-dependent
linearized Eliashberg equations with the
spin-fluctuation-mediated interaction potential $V(\vec q,\omega)$
from Eqs.(1), (3) and (5).
This gave a value of $g\!=\!0.64{\rm{eV}}$.
Using the same interaction potential,
with the same $g^2$ and $t_1$,
they then solved the full-zone $\vec k$- and $\omega$-dependent
Bethe-Salpeter equation in the normal-conducting state
to calculate the $T$-dependent resistivity
due to spin fluctuation scattering. This calculated model resistivity
was shown to be in reasonable agreement with the experimental data
on $YBa_2Cu_3O_{6.9}$\cite{ginsberg}.
Assumed input values and calculated results for the MP-II
model parameters are summarized in the last column of our Table I.

RULN, on the other hand, tried to base their parameter
values $t_1$ and $g^2$ entirely on
experimental DC and optical conductivity data,
rather than $T_c$ and LDA bandstructure results.
Their starting point are Drude model fits to the measured low-frequency
optical conductivity of $YBa_2Cu_3O_{6.7}$\cite{orenstein},
of the form
\begin{equation}
{\rm{Re}}\sigma(\omega)=\frac{\omega_{p,D}^2}{4\pi}
\frac{\tau_D}{1+\tau_D^2\omega^2}
\label{eq:7}
\end{equation}
parametrized by a Drude plasma frequency $\omega_{p,D}$
and a Drude relaxation time $\tau_D$\cite{footnote:Drude-lineshape}.
Having fixed $t\!\equiv\!t_2/t_1$ and the band-filling,
RULN determined their $t_1$ by equating
the experimental Drude plasma frequency $\omega_{p,D}$,
with the "bare" (non-interacting band) plasma frequency
\begin{equation}
\omega_{p,0}^2 = 2 N_L
\frac{4\pi e^2} {a_{\perp}} \int \frac{d^2k}{(2\pi)^2}
\delta(\epsilon_{\vec k}) v_x^2(\vec k)
\label{eq:8}
\end{equation}
where
\begin{equation}
v_x(\vec k)\equiv\frac{\partial}{\partial k_x}\epsilon_{\vec k},
\label{eq:9}
\end{equation}
$a_{\perp}$ is the $c$-axis lattice spacing
(perpendicular to the $CuO_2$ layers),
$N_L$ is the number of $CuO_2$ layers per 3D unit cell,
and the prefactor $2$ arises from the electron spin degeneracy.
Note here that, for fixed $t$, $a$, $a_{\perp}$ and band-filling,
$\omega_{p,0}^2$ is proportional to $N_L\!\times\!t_1$.
Setting $\omega_{p,0}$ equal to the experimental
$\omega_{p,D}\!=\!0.97(\pm0.05){\rm{eV}}$ for $YBCO$,
RULN thus obtained
$t_1\!=\!80(\pm10){\rm{meV}}$,
as listed in the column 1 of our Table I.
However, in this original estimate, RULN had
assumed an incorrect value of $N_L\!=\!1$,
corresponding to only one $CuO_2$ layer
per unit cell. With the correct number of $N_L\!=\!2$ $CuO_2$ layers
per unit cell in $YBCO$, one obtains a revised estimate for $t_1$
which is reduced from the original value
by a factor of 2, i.e. $t_1\!=\!40$meV,
since $\omega_{p,0}^2\!\propto\!N_L\!\times\!t_1$.
Our following discussion will be based on these
corrected values for $N_L$ and $t_1$, shown in column 2 of Table I.

Having thus fixed $t_1$, RULN then determined their
spin fluctuation coupling constant
$g^2$ by equating the experimentally determined
$T$-dependent Drude relaxation rate $1/\tau_D$, from Eq.(7),
with the calculated "bare" relaxation rate $1/\tau_0$, given by
\begin{equation}
1/\tau_0 \equiv 2\pi T \times \lambda_{tr,0}
= 2\pi T \times
       2\int_0^{+\infty}\frac{d\omega}{\omega}
       \; \alpha_{tr}^2F(\omega) \; \phi(\omega/T)
\label{eq:10}
\end{equation}
where $\phi(x)\!=\![(x/2)/{\rm{sinh}}(x/2)]^{2}$
in the variational Boltzmann\cite{allen}
and $\phi(x)=x/{\rm{sinh}}(x)$ in the
Fermi-surface-restricted Kubo
formalism\cite{ruln,fsk} for $x\!\equiv\!\omega/T$.
The transport spectral function
$\alpha_{tr}^2F(\omega)$, as given by Allen\cite{allen}, is
\begin{equation}
\alpha_{tr}^2F(\omega) =
\frac{a^2\int \frac{d^2k}{(2\pi)^2}\int \frac{d^2k'}{(2\pi)^2}
 \delta(\epsilon_{\vec k}) \delta(\epsilon_{\vec k'})
 (v_x(\vec k)-v_x(\vec k'))^2 {\rm{Im}} V(\vec k-\vec k',\omega+i0^+)}
{\pi \int \frac{d^2k}{(2\pi)^2}
\delta(\epsilon_{\vec k}) 2v_x^2(\vec k)}
\label{eq:11}
\end{equation}

{}From $\tau_0$ and $\omega_{p,0}$, the DC resistivity in this
formalism is calculated as
\begin{equation}
\rho=\frac{4\pi}{\omega_{p,0}^2\tau_0}
=\frac{8\pi^2T\lambda_{tr,0}}{\omega_{p,0}^2}
\label{eq:12}
\end{equation}
Notice for the following that,
as calculated from Eqs. (8-12),
for fixed $T$, $t$, $a$ and band-filling, $1/\tau_0$ and $\lambda_{tr,0}$
scale proportional to $g^2/t_1$, whereas $\rho$ scales proportional to
$g^2/t_1^2$.

By matching $1/\tau_0$, calculated from Eqs. (8-11),
in the
Fermi-surface-restricted Kubo formalism, to a conservative upper limit
on the then available experimental
$1/\tau_D$ data in the cuprates\cite{orenstein},
taken between about $120$K and $350$K,
RULN thus inferred an upper limit for the coupling
constant, $g^2\!=\!0.533{\rm{eV}}^2$,
based on their incorrect $t_1$ estimate of $80$meV,
as shown in the column 1 of
Table I.\cite{footnote:I-bar,footnote:tau_vs_lambda}
Using the corrected value of $t_1\!=\!40$meV, this
$g^2$ value also has to be corrected downward,
by a factor of 2, to
$g^2\!=\!0.267$eV$^2$, as shown in column 2 of Table I,
in order to maintain the same values of $1/\tau_0$ and $\rho$
as in the original RULN paper.
Note that, given the band parameters,
the foregoing procedure of determining $g^2$ is essentially
equivalent to adjusting $g^2$
so as to roughly fit the calculated resistivity from Eq.(12)
to the (upper limit) experimental resistivity
\begin{equation}
\rho\equiv 1/{\rm{Re}}\sigma(0)=\frac{4\pi}{\omega_{p,D}^2\tau_D} .
\label{eq:13}
\end{equation}

{}From standard transport theory\cite{allen,holstein}
it is well-known that neither $\omega_{p,0}$ and $\omega_{p,D}$
nor $\tau_0$ and $\tau_D$ are necessarily the same.
RULN's basic approach of simply equating
first $\omega_{p,0}\!=\!\omega_{p,D}$,
and then $\tau_0\!=\!\tau_D$,
to extract the bandwidth $8t_1$,
and, respectively, $g^2$ from the optical conductivity data,
is thus based on several simplifying assumptions. This
will be discussed in more detail in Section III of
this paper.

By solving the full-zone $\vec k$- and $\omega$-dependent
linearized Eliashberg equations for the RULN model,
with $g^2$ and $t_1$ values extracted from the optical
data, one then obtains an upper-limit $d$-wave $T_c$, from Fig. 1,
of about $11$K, as shown in columns 1 and 2 of Table I.
An important point to notice here is that RULN based their
parameter estimates on the then available\cite{ruln}
transport data on early YBCO samples\cite{orenstein} which had much larger
in-plane resistivities than the more
recently reported samples.\cite{ginsberg}
Consequently, RULN's model parameters
give a resistivity for {{}{their}} model\cite{ruln}
which is significantly larger than MP-II's $\rho$ values
for the MMP model or the recent experimental
$\rho$ data\cite{ginsberg}
on $YBa_2CuO_{6.9}$, used by MP-II
for comparison to their\cite{mp2} MMP model results.

Specifically,
using RULN's $\omega_{p,0}\!=\!0.97{\rm{eV}}$ and
their\cite{ruln} upper limit for
$\lambda_{tr,0}\!\equiv\!1/(2\pi T\tau_0)\!\sim\!0.52$
(approximately constant for $T\!>\!120{\rm{K}}$)
RULN have a relaxation rate
$1/\tau_0\!=\!2\pi\lambda_{tr,0}T\!\sim\!56{\rm{meV}}$
and a resistivity of about $450\mu\Omega$cm
at $T\!=\!200$K,
as shown in columns 1 and 2 of Table I
for the
Fermi-surface-restricted Kubo formalism.
By contrast, MP-II obtained about $64\mu\Omega$cm at $200$K,
from their solution of the full Bethe-Salpeter transport equation,
consistent with the in-plane-averaged resistivity of
$YBa_2Cu_3O_{6.9}$\cite{ginsberg}. From the width (HWHM)
of MP-II's ${\rm{Re}}\sigma(\omega)$-results
at $200$K, shown in Fig. 13 of MP-II,
we can also infer a rough estimate
for the {{}{calculated}} Drude relaxation rate $1/\tau_D$ and
Drude plasma frequency $\omega_{p,D}$ in the MMP
model\cite{mp2}, namely $1/\tau_D\!\sim\!12{\rm{meV}}$ and,
from Eq.(13),
$\omega_{p,D}\!=\!(4\pi/\rho\tau_D)^{1/2}\!\sim\!1.2{\rm{eV}}$,
as shown in the last column of Table I.\cite{footnote:Drude-lineshape}

Thus, for the respective model parameters used, MP-II and RULN
obtain or, respectively, assume
comparable values for the Drude plasma frequency,
$\omega_{p,D}\!\sim\!1$eV, consistent with the experimental
data. However, the Drude relaxation rate and the resistivity
of MP-II are {{}{smaller}} than those of RULN
by factors of about 5-7.
And yet, the $d$-wave $T_c$ of MP-II,
is {{}{larger}} than that of RULN
by a factor of about 9. We are thus called upon to explain
why $T_c$ in the RULN model comes out so much {{}{smaller}}
than in the MMP model, even for parameter
values giving the RULN model a substantially {{}{larger}}
resistivity than the MMP model.
\hbn
\hbn
\hbn
\section{ Mass renormalization, transport formalisms,
and Migdal parameters}
\label{sec:III}

In the present section, we will discuss in some detail the
major calculational differences, between RULN and MP-II,
in their determination of the bandwidth and in their
formalism for obtaining the resistivity.
We will elucidate the underlying assumptions
upon which the different approaches are based and point out
some of the potential short-comings of either approach.

First of all, regarding the bandwidth determination,
recall that the RULN estimate for
the bandwidth $8t_1\!\cong\!0.32{\rm{eV}}$,
corresponds to a non-interacting plasma frequency
$\omega_{p,0}$ which equals the observed Drude value for YBCO,
$\omega_{p,D}\!\cong\!0.97{\rm{eV}}$
(see columns 1 and 2, Table I).
This bandwidth is a factor of 6.25 smaller than the LDA-based
$8t_1\!=\!2.0{\rm{eV}}$ of MP-II. The latter
corresponds, by Eq.(8), to a non-interacting
plasma frequency $\omega^{(LDA)}_{p,0}\!\cong\!2.42{\rm{eV}}$,
at the RULN's assumed band-filling $n_h\!=\!1.18$ (see column 3, Table I).
It is presently not known whether the observed
optical mass enhancement
\begin{equation}
Z_D \equiv \Big( {\omega^{(LDA)}_{p,0}} \Big / {\omega_{p,D}}\Big)^2
\sim 6.25
\label{eq:14}
\end{equation}
is caused primarily by spin fluctuations or by non-spin-fluctuation
effects. It seems likely that both contributions are
present and non-negligible in the cuprates, that is
\begin{equation}
Z_D = Z_{nsf} \times Z_{sf}
\label{eq:15}
\end{equation}
with $Z_{sf}$ denoting a spin fluctuation and $Z_{nsf}$
a non-spin-fluctuation contribution, respectively.
In equating the observed Drude plasma frequency $\omega_{p,D}$
with {{}{their}} "bare" model plasma frequency $\omega_{p,0}$,
RULN thus implicitly made the following simplifying assumptions:

Firstly, they assumed that all the {{}{non-spin-fluctuation}}-induced mass
enhancement effects can be essentially
accounted for by using,
as {{}{input}} into the spin fluctuation calculation, an effective
bandwidth $8t_1$ and an effective spin fluctuation
coupling constant $g^2$ into which all non-spin-fluctuation
renormalization effects have been absorbed. Thus
\begin{eqnarray}
& &Z_{nsf} = \Big( {\omega^{(LDA)}_{p,0}} \Big / {\omega_{p,0}}\Big)^2
\nonumber\\
& &Z_{sf}  = \Big( {\omega_{p,0}} \Big / {\omega_{p,D}}\Big)^2.
\label{eq:16}
\end{eqnarray}
where $\omega_{p,0}$ is the effective "bare" plasma frequency,
corresponding to RULN's effective $t_1$ via Eq.(8).

Secondly, RULN assumed that the observed $Z_D$
is dominated by non-spin-fluctuation effects,
i.e. arises from local Coulomb
(and possibly other, such as electron-phonon)
interactions {{}{excluding}}
the effect of spin fluctuations, whereas the spin fluctuation
contribution is negligible, that is
\begin{eqnarray}
& &Z_D \cong Z_{nsf}
\nonumber\\
&|&Z_{sf}-1|  \ll  1
\label{eq:17}
\end{eqnarray}

{{}{Given}} these two assumptions,
the effective bandwidth and effective coupling constant
can be inferred directly from the experimental data
for $\omega_{p,D}$ and $1/\tau_D$, as was done by RULN.
Note however, that the term "bare," in the context of the RULN approach,
should then be understood to mean only "bare of renormalizations
due to spin fluctuations", and {{}{not}} "bare of {{}{all}} residual
interactions beyond LDA".

Recall here that, in conventional
Fermi-surface-restricted transport theory,\cite{allen,holstein}
the width (HWHM) of the Drude peak, $1/\tau_D$,
and its integrated spectral weight, $\omega_{p,D}^2$,
are related to $\tau_0$ and $\omega_{p0}$
as follows:
\begin{eqnarray}
\omega_{p,D}^2 & \cong & \omega_{p,0}^2/Z_{sf}
\nonumber\\
\tau_D & \cong & \tau_0 \times Z_{sf}.
\label{eq:18}
\end{eqnarray}
Thus, $1/\tau_D$ and $\omega_{p,D}^2$ are both renormalized
by $Z_{sf}$ in such a way that $Z_{sf}$
cancels out in the DC resistivity
\begin{equation}
\frac{4\pi}{\omega_{p,D}^2\tau_D} = \rho
=\frac{4\pi}{\omega_{p,0}^2\tau_0},
\label{eq:19}
\end{equation}
as already implied by Eqs.(12) and (13).
One can therefore indeed calculate the DC resistivity
$\rho\!\equiv\!1/\sigma(0)$
directly from $\omega_{p,0}^2$ and $1/\tau_0$,
as assumed by RULN in Eq.(12).\cite{ruln}
However, one may not equate $\omega_{p,0}$ with
the observed $\omega_{p,D}$, nor $\tau_0$ with $\tau_D$,
unless the conditions of
Eq.(17), i.e. $|Z_{sf}-1|\!\ll\!1$ are met, a point
on which we elaborate further below.

As far as spin- and non-spin-fluctuation
mass enhancements are concerned,
MP\cite{mp1,mp2} take the extreme opposite
point of view: They\cite{mp2} implicitly
assume that essentially {{}{all}} of the observed Drude mass enhancement
$Z_D$ is generated by the spin fluctuations and that all
non-spin-fluctuation effects are negligible, that is, in Eq.(15),
\begin{eqnarray}
& &Z_D  \cong  Z_{sf}
\nonumber\\
&|&Z_{nsf}-1|  \ll  1
\label{eq:20}
\end{eqnarray}
Given {{}{that}} assumption, one should compare the experimental data
{{}{only}} to the {{}{calculated}} model
Drude relaxation rate $1/\tau_D$ and calculated
Drude plasma frequency $\omega_{p,D}$,
which fully {{}{include}} all spin-fluctuation-induced
mass renormalization effects.
To estimate the values of $1/\tau_D$ and $\omega_{p,D}$
given in Table I for the MMP model,\cite{footnote:Drude-lineshape}
we have thus extracted $1/\tau_D$ directly
from the Drude peak width (HWHM) of the Bethe-Salpeter
solution for the optical conductivity $\sigma(\omega)$,
shown in Fig. 13 of MP-II.\cite{mp2} Using the $\rho$ value
(at $T\!=\!200$K) from Fig. 12 of MP-II, we have then obtained
$\omega_{p,D}$ from Eq.(13), i.e. by
$\omega_{p,D}\!=\!(4\pi/\rho\tau_D)^{1/2}$.
The resulting MMP model value for $\omega_{p,D}$, as discussed above,
is in rough agreement with the experimental data on YBCO.

Physically, both the RULN
and the MP treatment of the quasi-particle mass enhancement
raises some concerns. As far as the RULN approach\cite{ruln}
is concerned, their model assumption of simply absorbing
all non-spin-fluctuation effects into an effective $t_1$
and an effective $g^2$ clearly needs to be re-examined
and justified  within
the framework of a microscopic theory.
Also, for the $t_1$ and $g^2$ estimates given in column 2
of Table I, the spin fluctuation contribution $Z_{sf}$ to the overall
mass enhancement $Z_D$ may not really be negligible.
Namely, given the spin-fluctuation-mediated interaction
potential, Eq.(5), one can get a rough estimate of $Z_{sf}$
from the self-consistently calculated
single-particle self-energy $\Sigma(k,\omega+i0^+)$
due the spin fluctuations as follows:
\begin{equation}
Z_{sf} \sim
1 -
\Big\langle
\frac{\partial}{\partial \omega}
{\rm{Re}}\Sigma(\vec k,\omega+i0^+)|_{\omega=0}
\Big\rangle_{FS} \sim 1 + \lambda_Z > 1,
\label{eq:21}
\end{equation}
where $\langle ...\rangle_{FS}$ denotes a suitable
Fermi surface average and $\lambda_Z$ is the dimensionless
Eliashberg coupling parameter
\begin{equation}
\lambda_Z = 2 \int_{0}^{\infty} \frac{d\omega}{\omega}
\alpha_Z^2F(\omega),
\label{eq:22}
\end{equation}
obtained from the Eliashberg spectral function
of the spin-fluctuation-mediated interaction potential
\begin{equation}
\alpha_Z^2 F(\omega) =
\frac{a^2\int \frac{d^2k}{(2\pi)^2}\int \frac{d^2k'}{(2\pi)^2}
 \delta(\epsilon_{\vec k}) \delta(\epsilon_{\vec k'})
 {\rm{Im}}V(\vec k-\vec k',\omega+i0^+)}
{ \pi \int \frac{d^2k}{(2\pi)^2} \delta(\epsilon_{\vec k}) }
\label{eq:23}
\end{equation}
In the RULN model with the parameter values of columns 1 or 2, Table I,
and temperatures $T\!\lsim\!200$K one finds
$\lambda_Z\!\sim\!0.8-1.1$
and thus $Z_{sf}\!\sim\!1.8-2.1$ which is {{}{not}}
a negligible factor, contrary to
Eq.(17).\cite{footnote:T-dep.Z_sf}

To infer both $t_1$ and $g^2$ from the transport and optical data,
one should therefore really carry out a calculation for
the full dynamical conductivity $\sigma(\omega)$ in the RULN
model, including all spin-fluctuation-induced
mass-renormalization effects, and adjust both $t_1$ and $g^2$,
simultaneously, such that the calculated $1/\tau_D$,
from the Drude peak width, and the calculated $\omega_{p,D}$,
from $\omega_{p,D}\!=\!(4\pi/\rho\tau_D)^{1/2}$, Eq.(13),
both match the experimental values.
Such an approach would likely give substantially larger estimates
for both $t_1$ and $g^2$ if one assumes as input RULN's original
experimental $\omega_{p,D}$ and (upper-limit) $\rho$ values,
as given in columns 1 and 2 of our Table I.

However, if one uses the more recent, much lower $\rho$ data
for $YBCO$\cite{ginsberg}, the original RULN procedure
would give a much smaller $g^2$ and $\lambda_Z$ estimate,
with about the same $t_1$, as in column 2 of Table I.
The resulting $Z_{sf}$ would then indeed be
close to unity, i.e. the conditions of Eq.(17) would be satisfied.
So, based on the more recent, lower $\rho$ data, the original
RULN approach\cite{ruln} could actually be justified, at least as far
as consistency with the underlying assumptions about $Z_{sf}$
is concerned. It is obvious from Fig.1, that the resulting $T_c$,
based on these lower $\rho$ values, would be substantially {{}{below}} $11$K,
due to the smaller $g^2$ value.
So, if one were to use the RULN model susceptibility
together with the most recent transport data\cite{ginsberg}
for $YBCO$ as input
into the a fully self-consistent calculation of $\sigma(\omega)$,
including all spin-fluctuation-induced mass-renormalization effects,
one would ultimately find about the same $t_1$, but a substantially
smaller value for $g^2$ and $T_c$ than the original RULN estimates
stated in columns 1 and 2 of our Table I.

As far as the MP approach\cite{mp1,mp2} is concerned,
there is no good justification for assuming that
non-spin-fluctuation contributions to the mass enhancement
are negligible. As recently demonstrated
explicitly for the case of electron-phonon
coupling,\cite{hbs-chp}
non-spin-fluctuation contributions can significantly
suppress the $d$-wave $T_c$,
if $|Z_{nsf}-1|\!\sim\!{\cal{O}}(1)$.
In the MMP model, one is then forced to assume a substantially larger
spin fluctuation coupling $g^2$
than was used in MP-II, in order to maintain a $T_c\!\sim\!100$K.
This, in turn, will cause a larger model resistivity in the normal
state, above $T_c$. Thus, if substantial non-spin-fluctuation
contributions to the the mass enhancement are present,
the agreement between the MMP model results and the
experimental transport data becomes somewhat questionable.

A second major difference between MP\cite{mp1,mp2}
and RULN\cite{ruln} lies in their calculational approaches
for obtaining the resistivity.
Originally, MP had calculated the optical
conductivity $\sigma(\omega)$ diagrammatically
from the self-consistent
single-particle Greens function, in the simplest single-loop
approximation, without vertex corrections.\cite{mp1} Subsequently,
in MP-II,\cite{mp2} they tried to improve upon that approach
by including vertex corrections to the current correlation
function, at the level of a Bethe-Salpeter ladder approximation.\cite{mp2}
Their solution of this "full-zone" Bethe-Salpeter equation was done
numerically, without further approximations,
including the full $k$- and $\omega$-dependence
of the current vertex function.

RULN, on the other hand, relied on the variational Boltzmann
and
Fermi-surface-restricted Kubo formalisms\cite{allen,ruln,fsk}
outlined above. It is well-known\cite{allen,holstein}
that the semi-classical Boltzmann
equation\cite{allen} can be derived, under
certain simplifying conditions,
by an approximate Fermi-surface-restricted analysis
of the full-zone Bethe-Salpeter equations.
These simplifying conditions
are indeed well obeyed by conventional wide-band
electron-phonon systems.\cite{allen,holstein}
However, as discussed further below,
it is not clear whether these simplifying conditions
are also satisfied for the spin fluctuation models
under consideration here.

To test the various transport formalisms, we have calculated the
resistivity for the case of the MMP model,
using both of the Fermi-surface-restricted
approaches,\cite{allen,ruln,fsk}
and compared to the results obtained by MP-II\cite{mp2}
from numerical solutions of the full-zone Bethe-Salpeter
equation,  as shown in their\cite{mp2} Fig.12, with the
$T$-dependent spin fluctuation parameters given by Eq.(3).
The results of this comparison are shown in our Fig. 2.
As suggested in MP-II, we find that both
the variational Boltzmann and the
Fermi-surface-restricted Kubo formalism, as used by
RULN, do indeed overestimate the resistivity, compared to the
Bethe-Salpeter results. On the other hand, the overestimate,
being about a factor of 1.7 for the
Fermi-surface-restricted Kubo formalism at 200 K,
is by no means as large as was implied in MP-II.\cite{mp2}
For the variational Boltzmann formalism, the overestimate
is somewhat larger, about a factor of $2.1$ relative to the
Bethe-Salpeter results at $T\!=\!200$K.
Note here, that the $g^2$ estimate obtained by RULN
for {{}{their}} model\cite{ruln} was based on
resistivity calculations in the
Fermi-surface-restricted Kubo formalism.

Several causes could, potentially,
contribute to the differences between the full-zone
Bethe-Salpeter and the Fermi-surface-restricted
variational Boltzmann results shown in Fig. 2.
First of all, there is a numerical issue, having to do with the
method used in MP-II to calculate the optical
conductivity $\sigma(\omega)$ in the real-frequency
domain. In MP-II, the numerical solution of the
Bethe-Salpeter equation and the subsequent calculation of the
current-current correlation function
$C(\omega)\!\equiv\!i\omega\sigma(\omega)$ were actually
carried out in the Matsubara imaginary-frequency domain.
The Vidberg-Serene\cite{vidberg-serene} Pad\'e-approximant
technique was then used to analytically continue
the imaginary-frequency data $C(i\omega_m)$,
from imaginary frequencies $i\omega_m\!\equiv\!2\pi miT$
into the real-frequency correlation function
$C(\omega+i0^+)$, defined on the real-$\omega$ axis.
This analytical continuation procedure is mathematically
equivalent to an inverse Laplace transform
and well-known to be numerically highly unstable.
That is, even very small numerical errors in the
imaginary-frequency input data $C(i\omega_m)$,
for example due to finite cut-offs in the
Matsubara frequency summations, can produce
large errors in the real-frequency output
data $C(\omega+i0^+)$. This problem tends
to become particularly severe at higher temperatures
when the smallest non-zero Matsubara frequency off the
real axis, $i\omega_1$, becomes comparable or larger
in magnitude than the characteristic frequency scale one is
trying to resolve on the real-$\omega$ axis.
For example, in the MP-II calculation at $T\!=\!200$K,
$\omega_1\!=\!2\pi T\!=\!108$meV which is
more than 4 times larger than
the width of the real-frequency spectral feature, the Drude
peak width $2/\tau_D\!\cong\!24$meV, to be resolved.
It is not clear what level of numerical accuracy
was actually achieved in the analytical continuation
of MP-II.\cite{mp2}

Another potential source for discrepancies
between variational Boltzmann and full-zone
Bethe-Salpeter approach lies in the two-step approximation
which is being used to derive the former from the
latter.\cite{allen,holstein}
The first approximation step consists of restricting
the analysis of the full Bethe-Salpeter ladder equation
to the Fermi surface.\cite{allen,holstein}
This step reduces the Bethe-Salpeter equation
to the linearized semi-classical Boltzmann transport equation.
The second step is then to employ a simple variational ansatz
to obtain an approximate analytical solution to
the Boltzmann equation, which leads to our Eqs. (8-12).

The error from the second approximation step, the variational
ansatz, may become appreciable when the temperature falls below
the characteristic boson energy scale $\Omega$.
For example, in the case of acoustic phonon exchange,
the error is found to cause overestimates in $\rho$
as large as $30\%$ at temperatures of about $\frac{1}{8}$ of the
Debye phonon energy\cite{allen_var}.
In the case of the MMP model, the relevant boson energy
scale $\Omega$ is of the order of $100-150$meV,
as estimated e.g. by the spin fluctuation spectral
moments $\langle\omega^p\rangle_Z^{1/p}$ listed in
the last column of Table I and defined below. This is about a
factor of $6-9$ larger than $T$ at $T\!=\!200$K and
could therefore cause errors due the
variational ansatz to be of similar ($\sim\!30\%$) magnitude.
However, in the case of the RULN model, the respective
frequency moments (see Table I columns 1-4)
are about $2.5-3$ times smaller,
thus substantially reducing the error due to the variational
ansatz for that model.\cite{allen_var}

The first approximation step, reducing the Bethe-Salpeter
to the linearized semi-classical Boltzmann equation,
requires that the dimensionless Migdal parameter
\begin{equation}
s \equiv \lambda_Z \frac{\Omega}{\epsilon_F}
\label{eq:24}
\end{equation}
be small compared to unity.
Here, $\lambda_Z$ is the Eliashberg parameter,
from Eq.(22), $\epsilon_F$ is the Fermi energy,
measured from the nearest band edge, and
$\Omega$ is again the characteristic boson energy scale.
A reasonable order of magnitude for $\Omega$ can
be estimated e.g. by taking a low-order frequency moment of
the Eliashberg spectral function
$\alpha_Z^2F(\omega)$, Eq.(23), or of the transport
spectral function $\alpha_{tr}^2F(\omega)$, Eq.(11), that is, e.g.
\begin{equation}
\Omega\sim \langle\omega^p\rangle_Z^{1/p}
\label{eq:25}
\end{equation}
where
\begin{equation}
\langle\omega^p\rangle_Z
\equiv
{\int_0^{\infty}\frac{d\omega}{\omega}
 \alpha_Z^2 F(\omega) \omega^p}
\Bigg/
{\int_0^{\infty}\frac{d\omega}{\omega}
 \alpha_Z^2 F(\omega)}
\label{eq:26}
\end{equation}
The results for $\langle\omega^p\rangle_Z^{1/p}$ with $p\!=\!1$
and $2$ at $T\!=\!200$K are given for both models in Table I.
The moments $\langle\omega^p\rangle_{tr}^{1/p}$
of the transport spectral function
$\alpha_{tr}^2F$, from
\begin{equation}
\langle\omega^p\rangle_{tr}^{1/p}
\equiv
{\int_0^{\infty}\frac{d\omega}{\omega}
 \alpha_{tr}^2 F(\omega) \omega^p}
\Bigg/
{\int_0^{\infty}\frac{d\omega}{\omega}
 \alpha_{tr}^2 F(\omega)},
\label{eq:27}
\end{equation}
are not listed in Table I,
but are, in the RULN case, identical to the analogous moments
of $\alpha_Z^2F$
and, in the MMP case, within $1\%$ of
the analogous $\alpha_Z^2F$ moments.

In conventional wide-band metals, with transport scattering
dominated by phonons, we have $\lambda_Z\!\lsim\!1$, $\Omega$
of the order of the Debye energy and hence $s$ of order $10^{-2}$.
The Fermi-surface-restricted Boltzmann theory is thus expected
to work quite well in such systems. However, in the MMP model,
both $\alpha_Z^2F(\omega)$, from Eq.(23), and
$\alpha_{tr}^2F(\omega)$, from Eq.(11), exhibit a wide
peak, centered around $\omega_m\!\sim\!90-120$meV
(depending on $T$),
with substantial spectral weight extending
up to the cut-off energy of $\Omega_c\!=\!400$meV,
as shown for $T\!=\!200K$ in Figs. 3a and 3b.
With $\epsilon_F\!\sim\!4t_1\!\sim\!1.0$eV
(near $1\over2$-filling),
$\lambda_Z\!\sim\!1.6$ and $\Omega\!\sim\!100-150$meV
(from Eq.(25) with $p\!\sim\!1-2$),
the Migdal parameter $s$ at $200$K is about $0.16-0.24$ here.
Noticeable quantitative
discrepancies between the full-zone Bethe-Salpeter and
Fermi-surface-restricted Boltzmann theory
can therefore be expected.

For the case of the RULN model,
both $\alpha_{tr}^2 F(\omega)$ and $\alpha_Z^2F(\omega)$,
as shown in Figs. 3a and 3b, have a sharp peak around
$\omega_m\!\sim\!30-50$meV
and the spectral weight above $\omega_m$ decreases more rapidly
and is cut off altogether at $\Omega_c\!=\!100$meV.
This is reflected in a substantially smaller boson
energy $\Omega\!\sim\!30-50$meV, from Eq.(25) with $p\!\sim\!1-2$.
For the band and coupling parameters from column 2
of Table I, giving an Eliashberg parameter
$\lambda_Z\!\sim\!0.8$ and a Fermi
energy $\epsilon_F\!\sim\!4t_1\!=\!0.16$eV,
the RULN model Migdal parameter at $200$K,
$s\!\sim\!0.15-0.25$, is comparable to that of the
MMP model.

However, if one chooses $t_1$ and $g^2$
in the RULN model so as to give the same
bandwidth and the same resistivity as in MP-II,
one finds the parameter estimates given in column 3
of Table I and discussed further below.
For this parameter set, with $\epsilon_F\!\sim\!4t_1\!=\!1.0$eV
and $\lambda_Z\!\sim\!1.3$, the
Migdal parameter $s$ at $200$K is only of the order $0.04-0.07$,
i.e. noticably smaller than in the MMP model.
For this last RULN parameter set (column 3 of Table I),
we therefore expect the Fermi-surface-restricted
Boltzmann theory to provide a better approximation
to a full-zone Bethe-Salpeter calculation of the resistivity.

The overall magnitude of the Migdal $s$ parameter
in the MMP model raises serious concerns about the validity
of the entire spin fluctuation exchange
approach\cite{mp1,mp2} itself.
It is well-known,\cite{allen,holstein,textbooks}
that the validity of the Migdal-Eliashberg approximation
rests on the smallness of that {{}{same}} parameter $s$, Eq.(24),
which is also required to ensure the validity of the
Fermi surface-restricted
Boltzmann approach. In other words, the Migdal parameter
$s$ which controls the relative magnitude
of the discrepancy between full-zone
Bethe-Salpeter and Boltzmann theory
transport results, Fig. 2, also controls the relative magnitude
of those higher-order non-Migdal vertex corrections which have been
discarded, both in MP-II and in RULN,
in the single-particle self-energy,
in the particle-particle interaction kernel
of the linearized Eliashberg $T_c$-equations,
{{}{and}} in the particle-hole interaction
kernel of the Bethe-Salpeter equation.
One should therefore not be mislead
into believing that numerical solutions of the full-zone
Bethe-Salpeter equations necessarily constitute a
"better" transport theory than the
semi-classical Fermi-surface-restricted Boltzmann
approach.\cite{allen,holstein}
Either approach works well when $s$ is sufficiently
small compared to unity.
And either approach can be expected to give, at best,
{{}{only qualitatively}} correct results
when the $s$ parameter attains magnitudes as large as
estimated above.
Concerns regarding the applicability of the Migdal
approximation in spin fluctuation exchange models
have already been raised in several earlier and recent
studies.\cite{hertz-levin,millis-vert,bulut,schrieffer}
\hbn
\hbn
\hbn
\section{The effect of spectral weight distribution on $T_c$}
\label{sec:IV}

The differences in the assumed band parameters
and in the transport formalisms of RULN and MP-II certainly
contribute to differences in their respective $d$-wave $T_c$ values.
For the purpose of comparing the spin-fluctuation-mediated
pairing in the two models, it is therefore of interest to treat
the two models on an equal footing, as far as band parameters and transport
formalism are concerned. In order to demonstrate that
the differences in the $T_c$ values of the two models is
not primarily caused by the differences in band parameters
or transport formalism, we have thus carried out a $T_c$ calculation
for the RULN model with the following modified parameter set,
shown in column 3 of Table I:

Firstly, instead of $t_1\!=\!40$meV, we use the {{}{same}}
LDA-based value $t_1\!=\!250$meV
which was assumed by MP.\cite{mp1,mp2} Secondly, with
that $t_1$ value fixed, we adjust the RULN coupling
constant $g^2$ so that the resistivity, calculated
in {{}{both}} models at 200K
in the
Fermi-surface-restricted Kubo formalism,\cite{ruln,fsk}
is the {{}{same}}, namely
$\rho\!=\!109{\rm{\mu\Omega cm}}$.
Recall that in the
Fermi-surface-restricted Kubo formalism, Eqs.(8-12),
the resistivity scales proportional to
$g^2/t_1^2$, for fixed $t\!\equiv\!t_2/t_1$, band-filling and $T$.
Consequently, the modified $g^2$ value of $2.53{\rm{eV}}^2$,
thus obtained, in column 3, is about a factor of 9 larger than the
$g^2$ estimate in column 2, in spite of the fact
that the underlying $\rho$ value is almost a factor of 4 smaller
than that assumed in column 2.

The important point to notice now
is that the resulting $T_c\!=\!18.4$K for the modified RULN
parameter values, in column 3,
is increased only modestly over the original RULN
estimate of $11.2$K.
Thus, in comparing the column 3 results
for the RULN model to the MP-II results, in the last column
of Table I, we conclude that
the RULN $T_c$ value is about a factor of 5 lower
than the MP-II value, {{}{even if}}
we assume the {{}{same}} model
bandwidth, similar band-filling,
and coupling constants $g^2$
giving similar values for the DC resistivity,
calculated in {{}{both}} models,
by the {{}{same}} transport formalism.

In fact, as noted earlier\cite{rlsn-vhs} and clearly seen
in Fig. 1, $T_c$ as a function of $g^2$,
in the RULN model, does not increase indefinitely with increasing $g^2$,
but rather seems to approach a finite saturation value for $g^2\!\to\infty$.
This saturation $T_c$ does not exceed values of about $50$K
for the band parameters considered here.
Using transport data and/or LDA bandstructure results
to obtain actual estimates for $g^2$ and $t_1$
only serves to make this fundamental upper limit on
$T_c$ in the RULN model more stringent.
Thus, given the RULN spin fluctuation spectrum of Eq.(4), with the
spectral parameter values stated, it is simply not possible to generate,
within a conventional Eliashberg theory, a
$d$-wave $T_c$ of, say, $90$K or larger, {{}{regardless}}
of the choice of band parameters and coupling strength $g^2$.

Why then does the the RULN spin susceptibility give such a low $T_c$,
compared to the MMP model, even when its assumed
bandwidth and resistivity is comparable to that of MP-II ?
To answer this, we define a pairing spectral function
as follows
\begin{equation}
\alpha_{d}^2F(\omega) =
-\frac{a^2\int \frac{d^2k}{(2\pi)^2}\int \frac{d^2k'}{(2\pi)^2}
 \delta(\epsilon_{\vec k}) \delta(\epsilon_{\vec k'})
 \eta(\vec k) \eta(\vec k') {\rm{Im}}V(\vec k-\vec k',\omega+i0^+)}
{ \pi \int \frac{d^2k}{(2\pi)^2} \delta(\epsilon_{\vec k}) \eta^2(\vec k)}
\label{eq:28}
\end{equation}
where
\begin{equation}
\eta(\vec k) = cos(k_xa) - cos(k_ya)
\label{eq:29}
\end{equation}
is the basis function for nearest neighbor
$d_{x^2-y^2}$ pairing.
Note here that the sign in Eq.(28) is defined such that
a positive $\alpha_d^2F$ corresponds to attraction
in the $d$-wave pairing channel.
As shown in Fig. 3c, the spectral weight
distribution of $\alpha_{d}^2F$
is quite similar to that of $\alpha_Z^2F$ and
$\alpha_{tr}^2F$ in either model.
Note here that, again, in the MMP model $\alpha_{d}^2F$
has a very broad peak, at an $\omega_m\!\sim\!35-55$meV
(depending on $T$), with substantial
spectral weight extending out to the $400$meV cut-off.
And, again, in the RULN model the
spectral weight is much more narrowly peaked, at similar
peak position $\omega_m\!\sim\!30-50$meV,
and falls off rapidly well before the $100$meV cut-off.
We emphasize that this latter spectral weight distribution
is claimed to be in accord\cite{ruln,sizha} with the
currently available neutron scattering data on $YBCO$.

Recall now, from Eqs.(8-11), that the transport relaxation
rate $1/\tau_0$, and hence the resistivity $\rho$,
is most sensitive to the low-frequency part of $\alpha_{tr}^2F$,
up to $\omega\!\sim\!T$.
Thus, with their $\omega_{p,0}$ values being about equal,
the two models will give roughly the same resistivity
if their respective coupling constants $g^2$ have been adjusted
so that their $\alpha_{tr}^2F$ are roughly the same, in absolute
magnitude, at low frequencies $\omega\!\lsim\!T$.
However, given such coupling constant values,
the MMP model will then give a much larger superconducting $T_c$
than the RULN model, because of its substantially larger
spectral weight in $\alpha_{d}^2F$
at high frequencies. The basic physical reason
for this is that, unlike the normal state DC resistivity $\rho$,
the superconducting transition temperature $T_c$ is quite
sensitive to the high-frequency part of the boson spectrum.

To analyze this more quantitatively,
we consider the moments and the dimensionless
Eliashberg coupling parameters associated with
$\alpha_d^2F$,\cite{ad,allen-mitro}
\begin{equation}
\langle\omega^p\rangle_{d} =
{\int_0^{\infty} \frac{d\omega}{\omega}
                       \alpha^2_{d}F(\omega) \omega^p}
\Bigg/
{\int_0^{\infty} \frac{d\omega}{\omega}
                       \alpha^2_{d}F(\omega)}
\label{eq:30}
\end{equation}
and
\begin{equation}
\lambda_{d} = 2 \int_0^{\infty} \frac{d\omega}{\omega}
                             \alpha^2_{d}F(\omega).
\label{eq:31}
\end{equation}
Notice that all frequency moments defined here and above
($\langle\omega^p\rangle_Z^{1/p}$,
$\langle\omega^p\rangle_{tr}^{1/p}$
and $\langle\omega^p\rangle_{d}^{1/p}$) are independent of
the coupling strength $g^2$ and bandwidth $8t_1$,
assuming fixed $T$, band shape $t\!\equiv\!t_2/t_1$
and band-filling. All the $\lambda$ parameters
($\lambda_Z$, $\lambda_{tr,0}$, $\lambda_{d}$),
on the other hand, scale proportional to $g^2/t_1$,
at fixed $T$, $t$ and filling.

In approximate $T_c$-formulas of the
McMillan-Allen-Dynes variety\cite{ad,allen-mitro},
$T_c$ is given in terms of $\lambda_Z$, $\lambda_{d}$
and $\langle\omega^p\rangle_{d}^{1/p}$
by an expression of the form
\begin{equation}
T_c \cong \langle \omega^p\rangle_{d}^{1/p} F(\lambda_Z,\lambda_{d})
\label{eq:32}
\end{equation}
where $p$ can be weakly $\lambda$-dependent with
$0^+\!\le\!p\!\le\!2$.
$F$ is monotonically increasing as a function of $\lambda_d$
and monotonically decreasing as a function of $\lambda_Z$.
In the weak-coupling limit, $\lambda_{d},\lambda_Z\!\ll\!1$,
$F$ varies roughly exponentially with $\lambda_Z$ and $1/\lambda_d$,
that is
\begin{equation}
F(\lambda_Z,\lambda_{d}) \cong {\rm{const}}\times
\exp(-1/\lambda_{d}^{*}).
\label{eq:33}
\end{equation}
where
\begin{equation}
\lambda_{d}^{*} = \frac{\lambda_{d}} {1+\lambda_Z}.
\label{eq:34}
\end{equation}
Note however that, for the coupling strengths discussed
here, in both models $\lambda_d,\lambda_Z\!\gsim\!{\cal{O}}(1)$,
as shown in Table I.
So both the MMP and the RULN model are really in
a regime of intermediate, not weak, coupling.
In this intermediate-coupling regime, the variation of $T_c$ with $\lambda_Z$
and $1/\lambda_d$ is typically
less rapid than implied by Eq.(33).\cite{ad,allen-mitro}

The values of $\langle\omega\rangle_{d}$,
$\langle\omega^2\rangle_{d}^{1/2}$,
$\lambda_{d}$ and $\lambda_d^{*}$, evaluated at the respective
$d$-wave $T_c$'s, are shown in Table I.
In comparing the RULN results,
from e.g. column 3, to the MMP results, in the last column of Table I,
there are three key points to be noted:

The first point is that the
low-order frequency moments of $\alpha_d^2F(\omega)$
in the MMP model are again significantly
larger than those of the RULN model, by
a factor of about 2-3.
The second point is that also the pairing $\lambda$-parameter,
$\lambda_d$, in the MMP
model is about 2.2 times larger than the RULN value in column 3,
despite the fact that both models have about the same
$\lambda_{tr,0}$ at $200$K. The third point is that,
{{}{per}} $\lambda_d$, the MMP model
has a noticeably smaller mass-enhancement $\lambda$-parameter,
$\lambda_Z$, than
the RULN model, that is, the ratio $\lambda_Z/\lambda_d$
is about 0.88 in the MMP model, compared to 1.74 in the RULN model.
The last two points imply that $\lambda_d^{*}$ in the
RULN model is noticeably smaller
than in the MMP model.

In the McMillan-Allen-Dynes formalism,\cite{ad,allen-mitro}
each of the foregoing three differences
between the two models will tend to give the MMP model a
larger $T_c$ than the RULN model. The first two differences,
in $\langle\omega^p\rangle_{d}^{1/p}$ and $\lambda_d$, are an immediate
consequence of the substantial additional high-frequency
spectral weight in the MMP model, for $\omega\!\gsim\!50$meV,
as shown in Fig. 3c.
The third difference, in the $\lambda_Z/\lambda_d$-ratio,
is more subtle and arises from
differences in the detailed momentum dependences of
the two model susceptibilities.

To demonstrate more explicitly
that the difference in the high-frequency spectral weight
is indeed the primary cause for the differences
in the $T_c$'s of the two models, we now consider what happens
if we take, for example, the basic functional form of the RULN model,
but allow it to have a substantially larger boson frequency scale
than originally imposed by RULN on the basis of the
neutron scattering data.\cite{ruln,sizha} That is, we modify the
RULN spin susceptibility model by re-scaling
its energy spectrum, leading to a modified
susceptibility function
\begin{equation}
\chi_{RULN,f}(\vec q,\omega) \equiv \chi_{RULN}(\vec q,\omega/f)
\label{eq:35}
\end{equation}
with a boson energy scaling factor $f$
and $\chi_{RULN}$ from Eq.(4).
A re-scaled MMP model susceptibility $\chi_{MMP,f}$
could be defined analogously with the MMP suceptibility
$\chi_{MMP}$ from Eqs. (1) and (5).
Notice here that all frequency moments,
$\langle\omega^p\rangle_Z^{1/p}$, $\langle\omega^p\rangle_{tr}^{1/p}$
and $\langle\omega^p\rangle_{d}^{1/p}$,
scale linearly with $f$, whereas $\lambda_Z$ and $\lambda_{d}$
are independent of $f$, assuming $T$ and all other model parameters
are fixed. From full-zone Eliashberg calculations, we find that,
upon boson energy re-scaling at fixed $\lambda$ parameters, $T_c$
varies indeed roughly linearly
with $\langle\omega^p\rangle_{d}^{1/p}$ in either model, consistent with
the basic McMillan-Allen-Dynes approach, Eq.(32).
Thus, by increasing the boson energy scale, with $f\!>1$, at fixed $g^2$,
we can raise the $d$-wave $T_c$ in the RULN model to values
which are as large as those in the MMP model, or larger,
while at the same time {{}{lowering}} the model's
normal state resistivity, above $T_c$, from Eqs.(8-12).
Likewise, by lowering the overall boson energy scale, in either model,
with $f\!<\!1$, we would decrease $T_c$ and increase the
resistivity $\rho$. Notice that $\lambda_{tr,0}$ and
$\rho$, from Eqs. (8-12), {{}{decrease}} with increasing
boson energy scale, since increasing $f$, at fixed $T$,
actually {{}{reduces}} the relevant spectral weight
of $\alpha_{tr}^2F(\omega)$ at low frequencies
$\omega\!\lsim\!T$.

Suppose now we adjust both $g^2$ and $f$ in the re-scaled
RULN model, Eq.(35), such that {{}{both}}
its resistivity $\rho$ at $T\!=\!200$K (from the
the
Fermi-surface-restricted Kubo formalism, Eqs. (8-12), say)
{{}{and}} its $d$-wave $T_c$ (from full-zone
Eliashberg solutions) match those of the MMP model, as given
in the last column of Table I.
The results for this last RULN parameter set are shown in column 4
of Table I.
The required value for the scale factor is about $f\!=\!2.70$,
implying that, relative to column 3,
all the various RULN frequency moments are increased by
that factor. Also, the coupling constant $g^2$ and, hence all
$\lambda$ values, are substantially increased, by more than a
factor of 2, relative to those of column 3. This increase in $g^2$
is necessary in order to maintain a constant resistivity,
$\rho\!=\!109\mu\Omega$cm, i.e. in order
to compensate for the decreasing effect on $\rho$ caused
by raising $f$.

In comparing the RULN results from column 4
to the MMP results in the last column of Table I,
the crucial point to notice is that
now both models, having the same $T_c$ and $\rho$,
also have roughly, to within $20-30\%$, the same frequency moments.
The fact that now the (re-scaled) RULN model acquires
somewhat ($\sim\!20-30\%$) larger frequency moments than
the MMP model can be rationalized
by noting that the smaller $\lambda_d^{*}$ in the RULN
model has to be compensated for by a larger boson energy scale,
in order to get the same $T_c$ as in the MMP model.

{}From the foregoing results, it is clear though that
the spin fluctuation spectral weight distribution is {{}{the}}
central factor determining $T_c$ in either model and
that the differences in $T_c$ are not primarily caused
by differences in bandstructure parameters or transport formalism.
\hbn
\hbn
\hbn
\section{Summary}
\label{sec:V}

We can see now that the problem of whether high-$T_c$
superconductivity is explained by a
spin fluctuation model is closely tied to the question about which
type of dynamic susceptibility is more realistic, MMP or RULN.
One important feature of the MMP model susceptibility is
that its behavior is similar to an effective RPA
form which is known to fit Monte Carlo data on the single band Hubbard
model.\cite{bulut}  That is, the peak in ${\rm{Im}}\chi(\vec{q},\omega+i0^+)$
versus frequency is a strong function of momentum.
In fact, the momentum dependence is so strong that it would imply
the system to be very close to a magnetic instability, in contradiction
to experiment.  This problem is discussed further by Monthoux and
Pines.\cite{mp2}  By contrast, in the RULN model
the $\omega$ peak position is independent of momentum.
The latter behavior would seem unphysical from an effective RPA viewpoint,
but is claimed to be consistent with  presently available
neutron scattering data on $YBCO$.\cite{sizha}
On the other hand, it seems quite clear that substantial high frequency
spectral weight is present in the undoped insulating cuprate
parent compounds,
since their AF magnon spectra must extend up to several hundred
meV, assuming accepted estimates of the in-plane AF exchange constant
$J\!\gsim\!100$meV. It is therefore quite plausible
that additional high-frequency spectral weight may also
exist in heretofore unexplored energy regimes in the doped cuprates.

We thus feel that the crux of the matter lies in the neutron scattering data.
The MMP form is based on NMR data which is essentially a zero-frequency
measurement.  We would argue that it is somewhat dangerous to infer
the full frequency dependence of the susceptibility based on that data.
There is hope that sufficiently reliable neutron scattering
data, elucidating the complete momentum and frequency dependence,
will become available to resolve these issues.
We emphasize this since the significance and
interpretation of current neutron scattering data has been a very
controversial issue, especially in $YBCO$.

In this context the most recent neutron scattering data
by Yamada {\it et al.}\cite{yamada} and by Hayden {\it et al.}\cite{hayden}
on $La_{2-x}Sr_xCuO_4$ (LSCO, with doping $x$ of $14-15\%$)
are of considerable interest. These new data seem to imply\cite{hayden}
that in the doped samples, the high-energy
spectral weight (up to $\sim\!300$meV)
in ${\rm{Im}}\chi$ is substantially suppressed
relative to the undoped, insulating antiferromagnetic
parent ($x\!=\!0$) compound
and that substantial low-energy spectral weight ($\sim\!22$meV)
is being built up by doping.  Also, the width in momentum space is relatively
independent of frequency and implies a correlation length of order $a$.
This observation would be quite consistent with an RULN picture.
On the other hand, low frequency data\cite{mason} show incommensurate peaks
with a correlation length of 6.7 $a$.  This rapid change of width in momentum
space from low to high frequencies would be consistent with an MMP picture.
It thus appears at this point that
the new LSCO data are, in some sense, intermediate between
the RULN and the MMP models.  As Hayden {\it et al.} point out,\cite{hayden}
their data are consistent with a broadened magnon dispersion, the functional
form of which does not look like either model considered here.

On the theoretical side, we caution against
over-interpreting the quantitative significance
of results obtained from phenomenological
spin fluctuation exchange models of the type discussed here.
In order to produce a low resistivity and a high $T_c$,
the typically required magnitude of the Migdal parameter $s$
in these models is sufficiently large to render
the whole Migdal-Eliashberg approach invalid or, at the very least,
makes it a rather questionable foundation for a quantitative theory.
We also emphasize the often overlooked point
that the conditions of applicability for the
Fermi-surface-restricted Boltzmann transport theory are essentially the
same as the conditions of validity for the Migdal approximation.
Thus, as far as different transport formalisms are concerned,
the Fermi-surface-restricted Boltzmann formalism
is in principle "no worse" than the full-zone Bethe-Salpeter theory
and, along with the Migdal approximation,
either approach becomes quantitatively unreliable when the
Migdal parameter $s$ is as large as we have found it for the
spin fluctuation models considered here.

Finally, the required, rather large boson energy
scales, with $\Omega/\epsilon_F\!\gsim\!0.1-0.2$,
imply that extended (and, especially, nearest neighbor) Coulomb
repulsions will substantially suppress the $d$-wave $T_c$
in the current AF spin fluctuation exchange models,
since the conventional mechanism of reducing the effective
Coulomb interaction strength via retardation\cite{allen-mitro}
becomes largely
inoperative. Calculations based on current spin fluctuation
models have so far neglected extended Coulomb interactions entirely
and may thus seriously overestimate the maximum
achievable $T_c$.\cite{hbs-unpub}

\acknowledgements

The authors would like to thank P. B. Allen,
P. Monthoux and R. J. Radtke for
several helpful discussions.
This work was supported by the National Science Foundation
(DMR 91-20000)
through the Science and Technology Center for Superconductivity (H.B.S.)
and by the U.~S.~Department of Energy, Basic Energy Sciences,
under Contract \#W-31-109-ENG-38 (M.R.N.).  H.B.S. would also like to
acknowledge support from the National Science Foundation
under grant No. DMR-9215123 and computing support from
the University of Georgia and from the
National Center for Supercomputing Applications
at the University of Illinois.
M.R.N. acknowledges support from the Aspen Center for Physics
where part of this manuscript was completed.

\begin{table}\caption{Comparison
of the RULN model, Eq.(4),
and the MMP model, Eqs.(1) and (3).
Abbreviations $FK$, $VB$, and $BS$ indicate transport results obtained
at $T\!=\!200$K
by the Fermi-surface-restricted Kubo and
variational Boltzmann formalism,  Eq.(10),
and by solution of the full-zone Bethe-Salpeter equation, from
Ref. [11], respectively.}
\begin{tabular}{llddddd}
Model:  &&            RULN &  RULN &  RULN &  RULN &    MMP \\
Column: &&                1&      2&      3&      4&      5 \\
\tableline
Model input parameters:  \\
\tableline
$t_1$ [meV]&&            80&     40&    250&    250&    250 \\
$g^2$ [eV$^2$]&&      0.533&  0.267&   2.53&   6.82&  0.410  \\
%
%
$f$&&                  1.00&   1.00&   1.00&   2.70&   1.00 \\
$N_L$&&                   1&      2&      2&      2&      2 \\
$n_h$&&                1.18&   1.18&   1.18&   1.18&   1.25 \\
\tableline
$d$-Wave $T_c$ and transport: \\
\tableline
$T_c$ [K]&&            10.6&   11.2&   18.4&   99.5&   99.5 \\
$\rho$ [$\mu\Omega$cm]&
FK&                    447.&   447.&   109.&   109.&    109. \\
&VB&                   556.&   556.&   136.&   150.&    132. \\
&BS&                     --&     --&     --&     --&     64. \\
$\omega_{p,0}$ [eV]&&  0.97&   0.97&   2.42&   2.42&   2.30 \\
$\omega_{p,D}$ [eV]&
BS&                      --&     --&     --&     --&   1.2  \\
$1/\tau_0$ [meV]&
FK&                    57. &   57. &   86. &   86. &   78.  \\
&VB&                   71. &   71. &  107. &  119. &   94.  \\
&BS&                     --&     --&     --&     --&   46.  \\
$1/\tau_D$ [meV]&
BS&                      --&     --&     --&     --&   12.  \\
$\lambda_{tr,0}$&
FK&                   0.52 &  0.52 &  0.79 &  0.79 &  0.72  \\
&VB&                  0.65 &  0.65 &  0.99 &  1.09 &  0.87  \\
&BS&                     --&     --&     --&     --&  0.42  \\
\tableline
Eliashberg $\lambda$ parameters: \\
\tableline
$\lambda_Z$&
  at $T_c$ &          1.07 &  1.07 &  1.64 &  4.11 &  1.78  \\
& at $200$K &         0.83 &  0.83 &  1.27 &  3.41 &  1.62  \\
$\lambda_d$& at $T_c$&
                      0.62 &  0.62 &  0.95 &  2.37 &  2.03  \\
$\lambda_d^{*}$& at $T_c$&
                      0.30 &  0.30 &  0.36 &  0.46 &  0.73  \\
\tableline
Frequency moments: \\
\tableline
$\langle\omega\rangle_Z$ [meV]&
  at $T_c$ &          32.5 &  32.5 &  32.8 &  98.  &  98.   \\
& at $200$K &         40.3 &  40.3 &  40.3 & 109.  & 105.   \\
$\langle\omega^2\rangle_Z^{1/2}$ [meV]&
  at $T_c$ &          36.9 &  36.9 &  37.4 & 115.  & 141.   \\
& at $200$K &         47.4 &  47.4 &  47.4 & 128.  & 146.   \\
$\langle\omega\rangle_d$ [meV]& at $T_c$ &
                      32.5 &  32.5 &  32.8 &  98.  &  71.   \\
$\langle\omega^2\rangle_d^{1/2}$ [meV]& at $T_c$ &
                      36.9 &  36.9 &  37.4 & 115.  & 109.   \\
\end{tabular}
\end{table}

\begin{figure}
\caption{$d$-wave superconducting transition temperature $T_c$
versus scaled coupling strength $g^2/t_1$ for the RULN model
interaction potential $V(q,\omega)\!=\!g^2\chi_{RULN}(q,\omega)$,
Eqs.(4) and (5), with $t\!\equiv\!t_2/t_1\!=\!0.45$ and
different values of $t_1$,
as indicated in the figure. $T_c$ was calculated by full-zone
solutions of the linearized Eliashberg equations on a $k$-mesh
corresponding to $64\!\times\!64$ $k$-points
covering the full 1st Brillouin zone,
with fermion Matsubara frequencies $\nu_m\!\equiv\!(2m-1)\pi T$
up to $m\!=\!1024$, using FFT techniques,
as described in Refs.[9-11,14-15].
\label{fig1}}
\end{figure}

\begin{figure}
\caption{Resistivity $\rho$ versus temperature $T$
for the MMP model, obtained
from the variational Boltzmann ($VB$, dot-dashed line)
and
from the Fermi-surface-restricted Kubo formalisms ($FK$, dashed line)
Eqs.(8-12).
For comparison, we also show
the results of Monthoux and Pines, MP-II, Ref.[11],
obtained from full-zone solutions of the
Bethe-Salpeter transport equation ($BS$, full line)
for identical model parameters, as given in the last
column of Table I, with $T$-dependent
$\omega_{sf}(T)$, $\chi_Q(T)$ and $\xi(T)$ given by Eq.(3).
\label{fig2}}
\end{figure}

\begin{figure}
\caption{
(a) $\alpha_{tr}^2F(\omega)$, Eq.(11),
(b) $\alpha_Z^2F(\omega)$, Eq.(23), and
(c) $\alpha_{d}^2F(\omega)$, Eq.(28),
versus frequency $\omega$ for the MMP model, Eq.(1),
and for the RULN model, Eq.(4),
at different temperatures $T$.
The RULN model parameters used are those from column 3
of Table I.
The MMP model parameters used are those from the
last column of Table I with $T$-dependent
$\omega_{sf}(T)$, $\chi_Q(T)$ and $\xi(T)$ given by Eq.(3).
The 3 different temperatures include
the $d$-wave superconducting $T_c$ values of the
RULN model, $T_{c,RULN}\!=\!18.4$K, and of the MMP
model, $T_{c,MMP}\!=\!99.5$K, for the respective model
parameters, as given in Table I, columns 3 and 5, respectively.
\label{fig3}}
\end{figure}


\begin{references}
\bibitem{hbs-unpub}
By including extended Coulomb repulsion terms into our full-zone
Eliashberg equations for the MMP model, Refs. [7-11],
we have shown that even a small nearest neighbor Coulomb
repulsion, of strength $V_1\!=\!200$meV, is sufficient to suppress
the $d$-wave $T_c$ by about a factor of $2$.
\bibitem{thanks} The authors would like to
thank P. Monthoux for clarifying discussions
concerning this point.
\bibitem{footnote:Drude-lineshape} Both the experimental
data and calculated  model results for $\sigma(\omega)$
may already deviate noticeably from a simple Drude law, Eq.(7),
for photon frequencies $\omega\!\gsim\!1/\tau_D$. To get at least
a meaningful order-of-magnitude estimate for a "Drude" plasma
frequency $\omega_{p,D}$, we have thus parametrized the
zero-frequency peak in $\sigma(\omega)$ in terms of its
peak height $\sigma(0)\!\equiv\!1/\rho$ and in terms of its
peak width, $1/\tau_D$, defined here as the half-width
at half-maximum (HWHM). We then use Eq.(13) as the defining
relation for $\omega_{p,D}$, that is
$\omega_{p,D}\!\equiv\!(4\pi/\rho\tau_D)^{1/2}$.
\bibitem{footnote:I-bar} In their original paper, RULN, Ref.[12],
stated their coupling strength in terms of a dimensionless
coupling parameter ${\bar{I}}^2$ which, in our present
notation, is given by ${\bar{I}}^2\!=\!g^2N_F C /\pi$
where $N_F$ is the non-interacting density of states
at the Fermi level. For a hole concentration $n_h\!=\!1.18$
holes per planar $Cu$-site (i.e. a hole doping concentration
of $x\!=\!18\%$) and $t\!\equiv\!t_2/t_1=0.45$, we have $N_F\!=\!0.2481/t_1$
and hence a conversion factor $N_F C/\pi=0.01500{\rm{eV}}^{-1}/t_1$.
Thus, for an assumed $t_1\!=\!80$meV, the original upper limit value of
${\bar{I}}^2\!=\!0.10$, proposed by RULN, corresponds to
a value of $g^2\!=\!0.533{\rm{eV}}^2$.
\bibitem{footnote:tau_vs_lambda} In RULN, Ref. [12], the
transport relaxation time results where actually expressed in terms
of the dimensionless transport $\lambda$ parameter,
$\lambda_{tr,0}\!\equiv\!(2\pi T\tau_0)^{-1}$,
from Eq.(10), which turns out to be approximately $T$-independent
in the RULN model for  $T\!\gsim\!200$K. RULN adjusted their
coupling strength $g^2$ (or $\bar{I}^2$, see previous footnote)
so as to match the calculated $\lambda_{tr,0}$ to the
measured, approximately $T$-independent
$\lambda_{tr,D}\!\equiv\!(2\pi T\tau_D)^{-1}$.
This procedure is of course entirely equivalent to adjusting $g^2$
so that $\tau_0$ becomes equal to $\tau_D$.
\end{references}
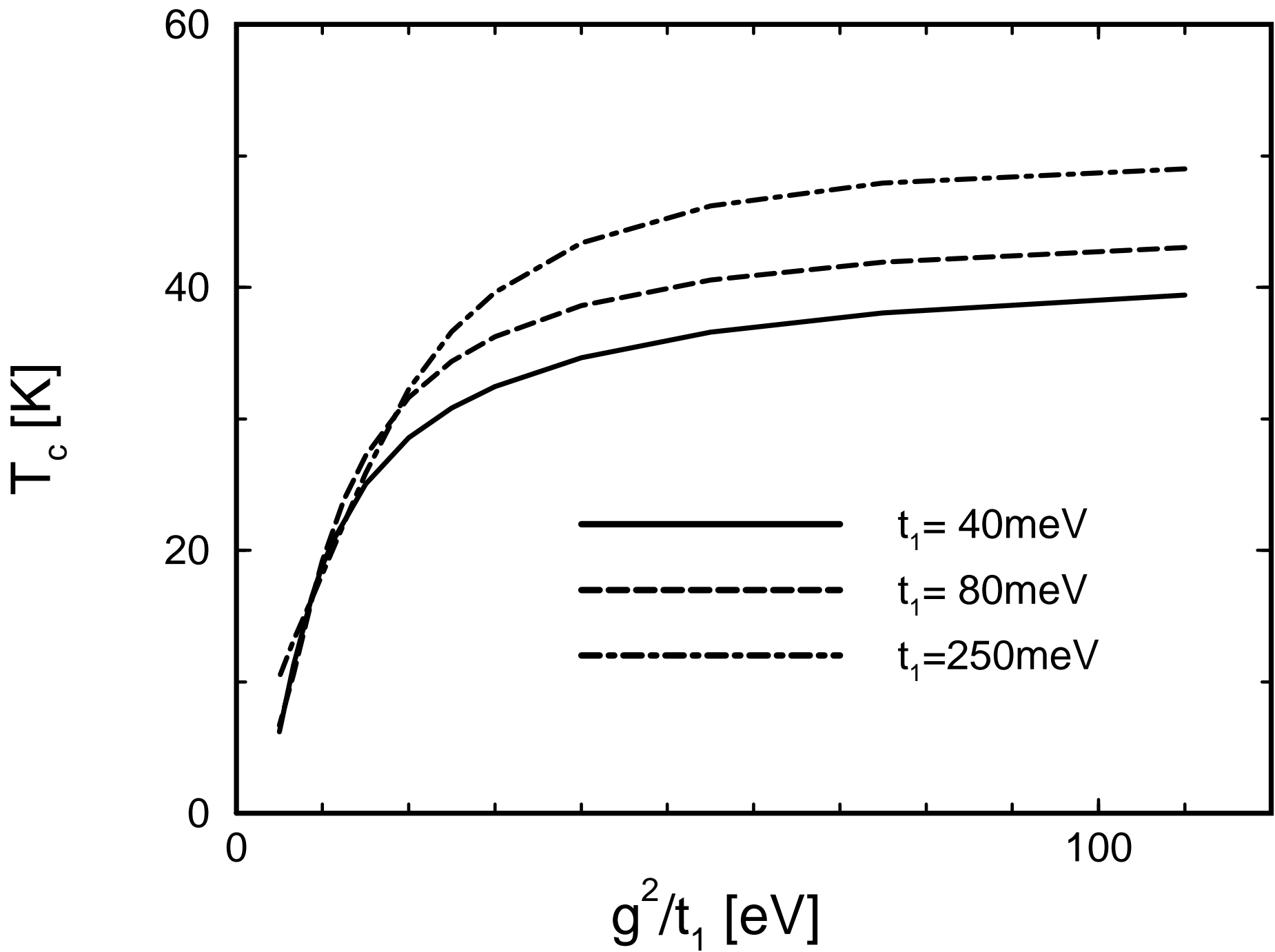
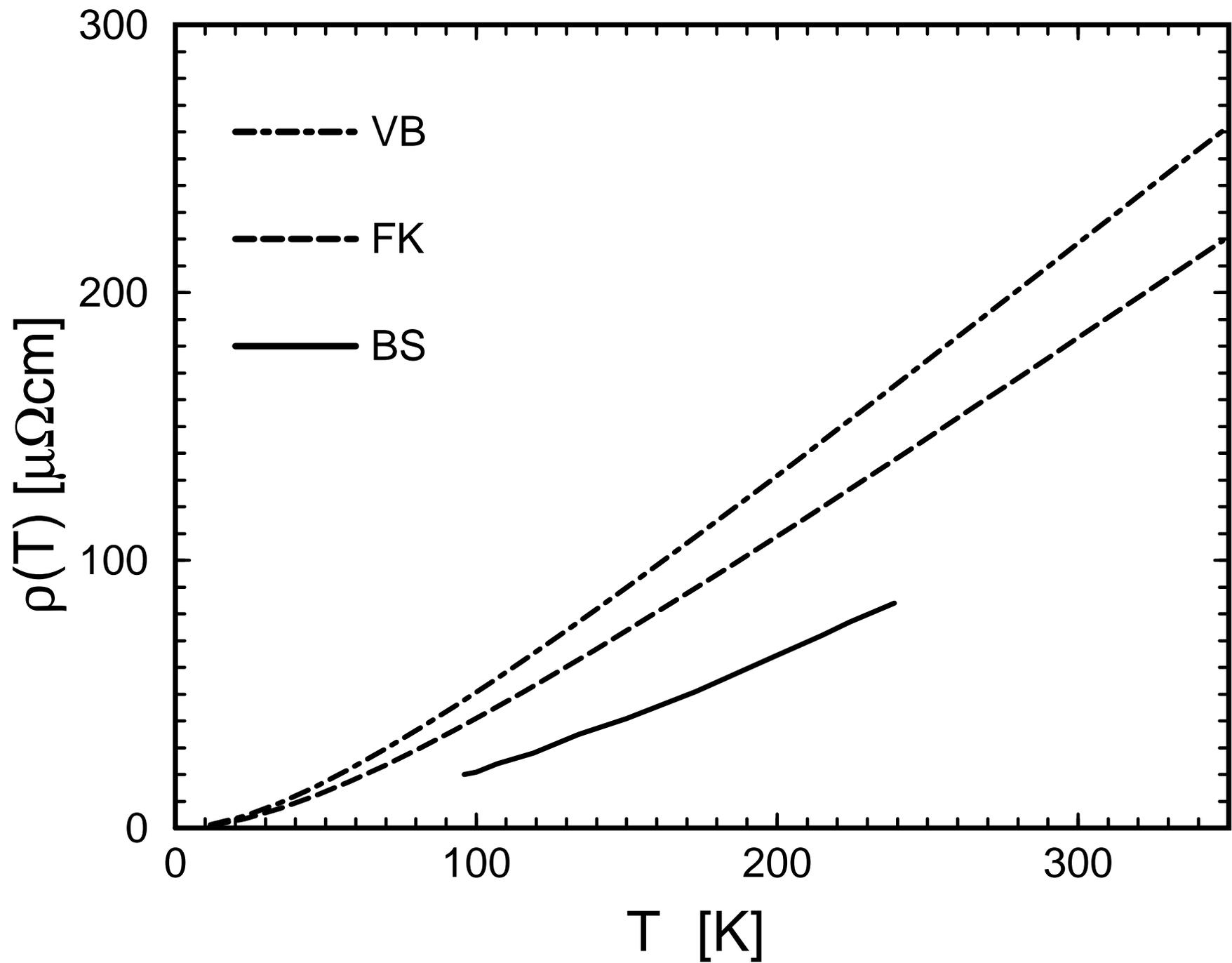
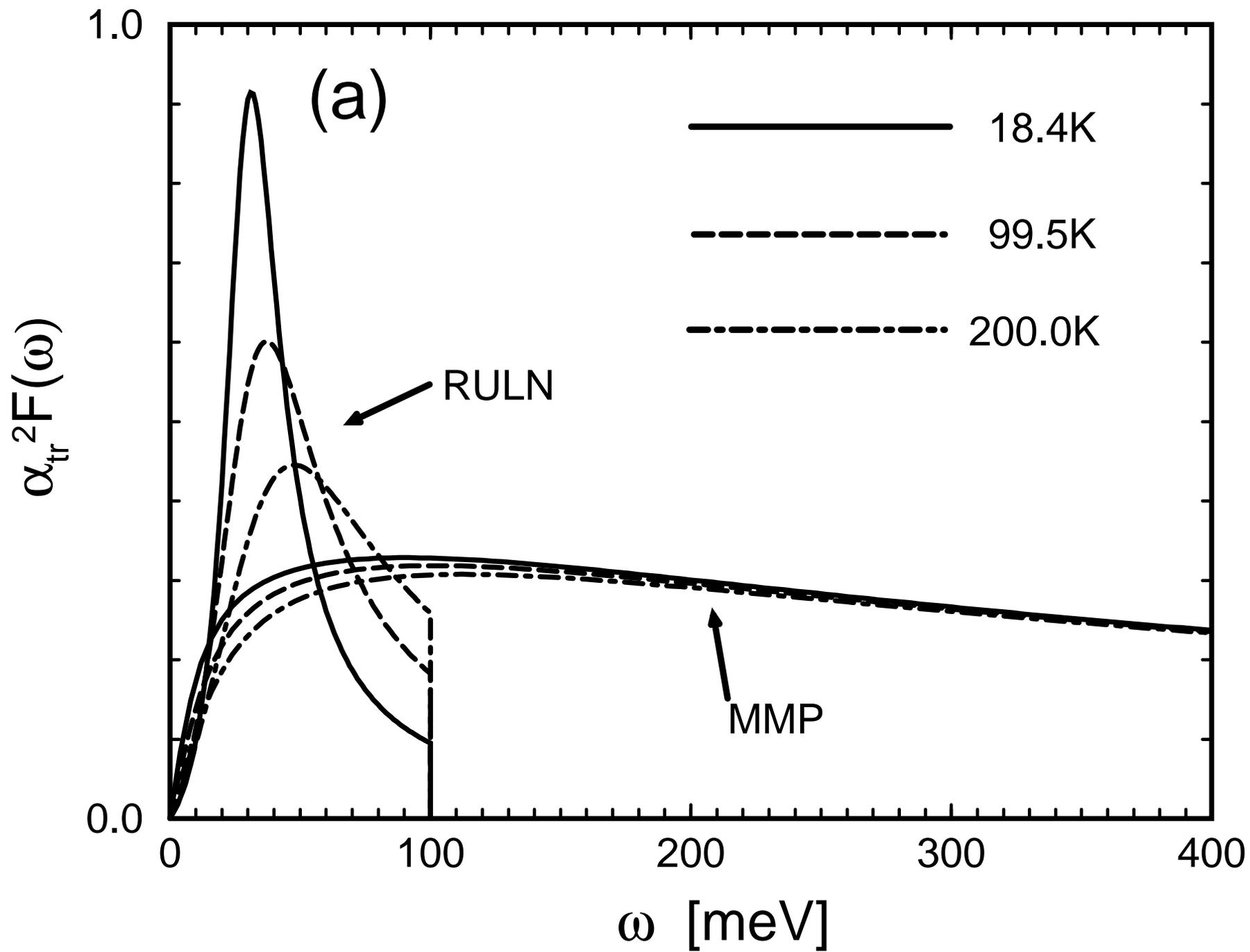
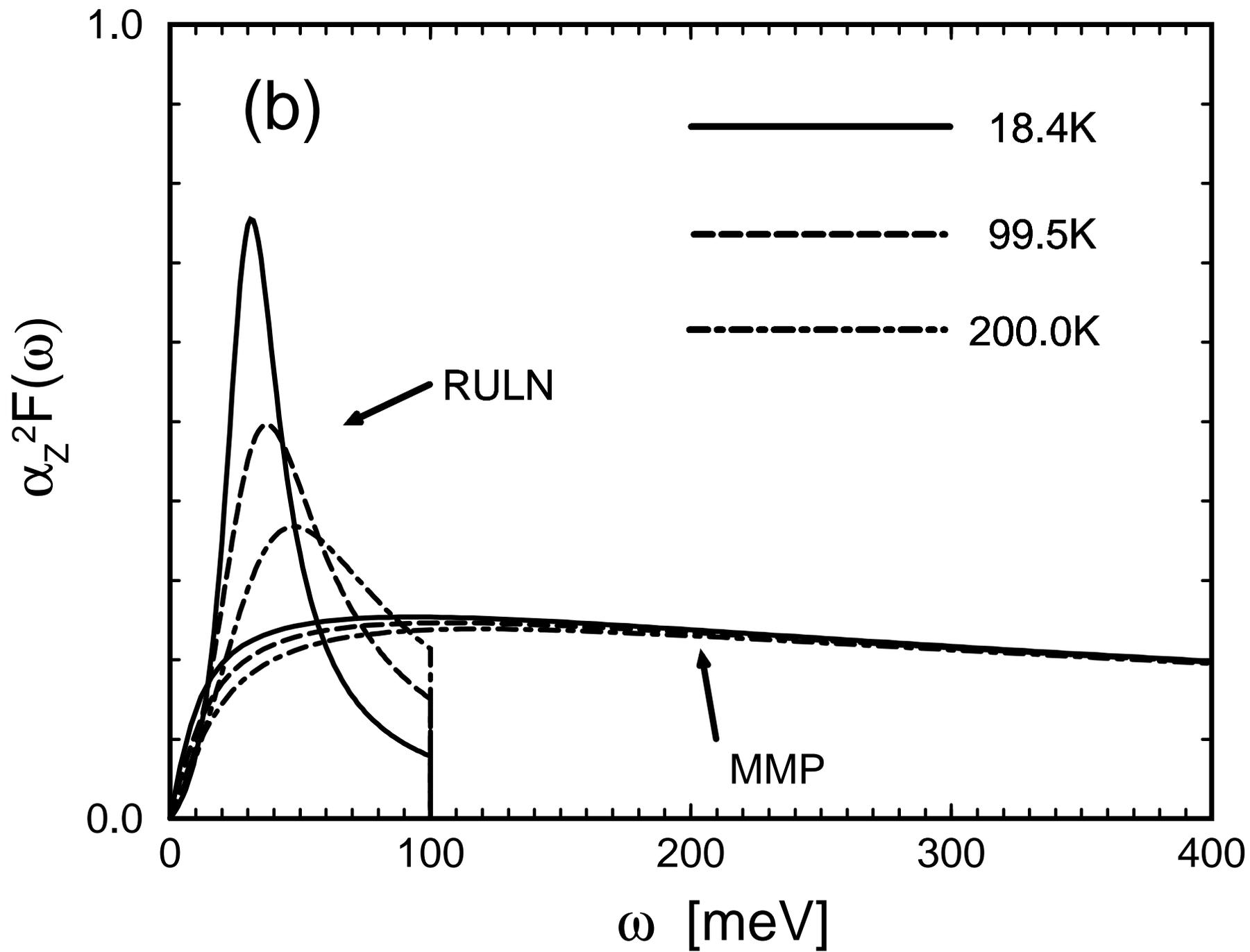
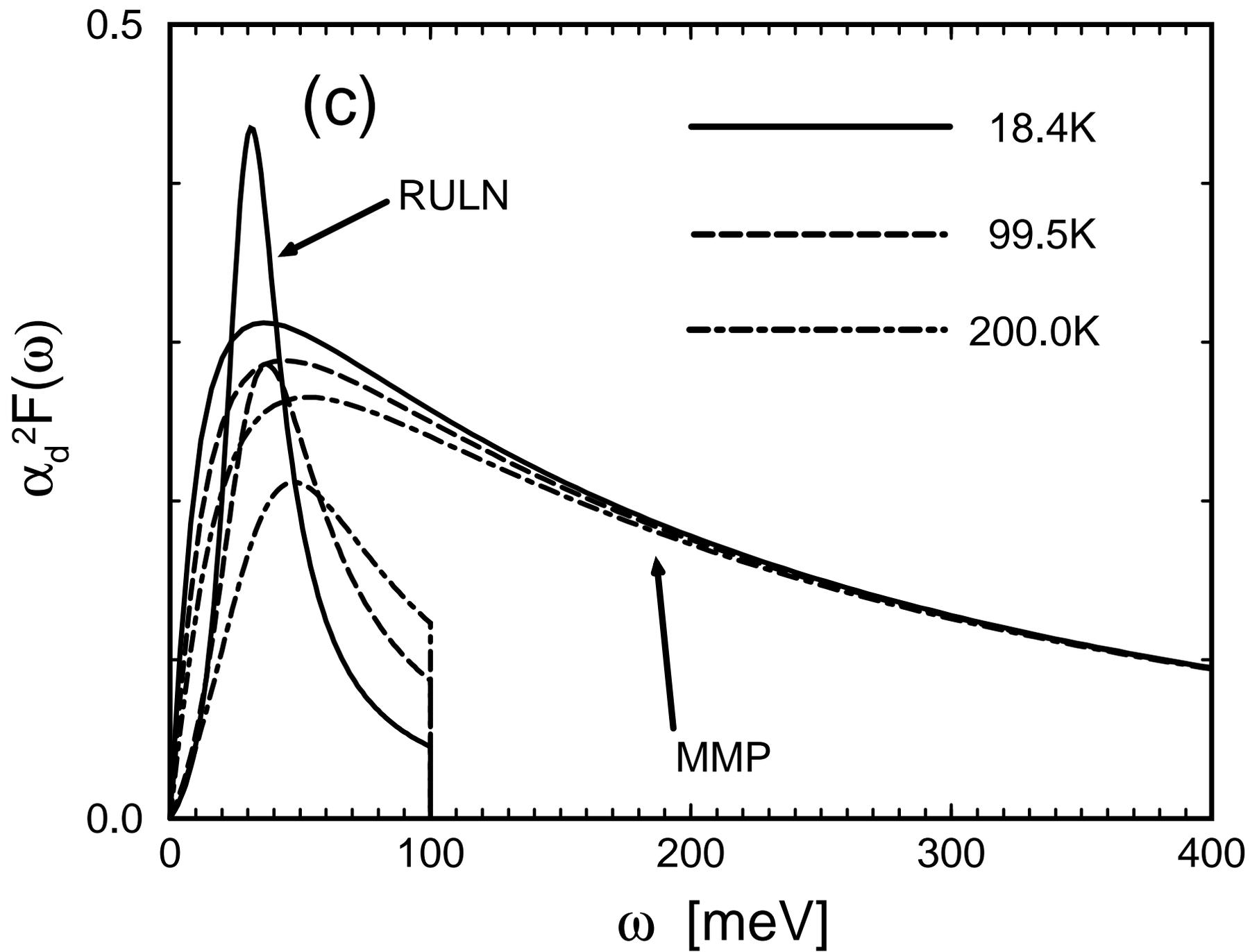
\end{document}